\documentclass[10pt,a4paper,final]{iopart}
\usepackage{iopams}
\usepackage{esint}
\usepackage{mathrsfs}
\usepackage{bm}
\usepackage{color}
\usepackage{graphicx}
\usepackage{subfig}
\usepackage{hyperref}
\usepackage{lineno}
\begin{document}
%\title{Fully Nonlinear Gyrokinetic Equations for Magnetized Plasmas}
%%\author{Hao-Tian Chen}\email[]{Email: haotianchen@zju.edu.cn}
%%\affiliation{Institute for Fusion Theory and Simulation and Department of Physics, Zhejiang University, Hangzhou, 310027, People's Republic of China}
%\author{Liu Chen}\email[]{Email (Corresponding author):liuchen@zju.edu.cn}
%\affiliation{Institute for Fusion Theory and Simulation and Department of Physics, Zhejiang University, Hangzhou, 310027, People's Republic of China}
%\affiliation{Department of Physics and Astronomy, University of California, Irvine, California 92697, USA}
%\author{Fulvio Zonca}
%\affiliation{ENEA, Fusion and Nuclear Safety Department, C. R. Frascati, Via E. Fermi 45, 00044 Frascati (Roma), Italy}
%\affiliation{Institute for Fusion Theory and Simulation and Department of Physics, Zhejiang University, Hangzhou, 310027, People's Republic of China}
%\vskip 0.25cm

\title[]{A Gyrokinetic Simulation Model for Low Frequency Electromagnetic Fluctuations in Magnetized Plasmas}
\author{Liu Chen$^{1,2}$, Haotian Chen$^{3}$\footnote{Author to whom any correspondence should be addressed.}, Fulvio Zonca$^{4,1}$ and Yu Lin$^{5}$}
\address{$^{1}$Institute for Fusion Theory and Simulation and Department of Physics, Zhejiang University, Hangzhou, 310027, People's Republic of China}
\address{$^{2}$Department of Physics and Astronomy, University of California, Irvine, California 92697, USA}
\address{$^{3}$Institute of Space Science and Technology, Nanchang University, Nanchang, 330031, People's Republic of China}
\address{$^{4}$Center for Nonlinear Plasma Science and C.R. ENEA Frascati - C.P. 65, 00044 Frascati, Italy}
\address{$^{5}$Physics Department, Auburn University, Auburn, AL 36849, USA}
%\eads{\mailto{haotianchen@ncu.edu.cn} }
\vskip 0.25cm

\date{\today}

\begin{abstract}
	We present a new model for simulating the electromagnetic fluctuations with frequencies much lower than the ion cyclotron frequency in plasmas confined in general magnetic configurations. 
	This novel model (termed as GK-E\&B) employs nonlinear gyrokinetic equations formulated in terms of electromagnetic fields along with momentum balance equations for solving fields.
	It, thus, not only includes kinetic effects, such as wave-particle interaction and microscopic (ion Larmor radius  scale) physics; but also is computationally more efficient than the conventional formulation described in terms of potentials. 
	As a benchmark, we perform linear as well as nonlinear simulations of the kinetic Alfv\'{e}n wave;  demonstrating physics in agreement with the analytical theories.
\end{abstract}


\maketitle
%\linenumbers
%\ioptwocol

\section{Introduction}
%\label{sec:may:01:16:01}
Electromagnetic fluctuations with frequencies lower than the ion cyclotron frequencies are prevalent in plasmas, existing in nature; e.g., solar and magnetospheric plasmas, and laboratories.
These low-frequency fluctuations are known to play important dynamic roles in the heating, acceleration and transports of plasmas.
For example, Alfv\'{e}n waves have been proposed as the mechanism responsible for the charged particle heating in the solar corona \cite{osterbrock, wentzel74, wentzel76,ionson78,ionson82}.
On the other hand, Alfv\'{e}n-wave instabilities can also be readily excited by energetic particles, including alpha particles,  in magnetic fusion devices, and thus, could be detrimental to the goal of magnetic fusion energy \cite{chen16}.
Due to the intrinsically complicated magnetic field geometries and plasma inhomogeneities, dynamics of waves and charged particles in low-frequency electromagnetic fluctuations are rather complex and involve, in general, disparate spatial as well as temporal scales. 
To be more specific, the existence of the shear Alfv\'{e}n wave continuous spectrum \cite{grad69} could lead to the appearance of kinetic Alfv\'{e}n wave at the microscopic ion-Larmor-radius ($\rho_{i}$) scale, which is, at least, $\mathcal{O}(10^{-3})$ smaller than the system-size macroscopic scale \cite{hasegawa75,hasegawa76, chen16}.
Similarly, as an example, the wave periods of Alfv\'{e}n-wave instabilities in magnetic fusion experiments are, typically, $\mathcal{O}(10^{-2})$ shorter  than the  nonlinear time scales \cite{zonca151, zonca152}. These, meanwhile, are comparable to the inverse of the instability growth rates and are $\mathcal{O}(10^{-1})$ shorter than those of resistive/collisionless tearing instabilities \cite{furth63, furth73, drake77}.
The complexities of nonlinear wave and charged-particle dynamics with at least $\mathcal{O}(10^{3})$ separation of spatial and temporal scales in nonuniform plasmas with complicated magnetic field geometries, thus, naturally demand the employment of numerical simulations as a powerful tool for understanding the observations, extracting the underlying physics mechanisms, and predicting the future performances \cite{chen16, falessi}.
Most of the existing simulation models are based on the so-called magnetohydrodynamic fluid approximation, which can not adequately address the kinetic physics such as enhanced space-charge separation at the microscopic $\rho_{i}$ scale and wave-particle interactions. 
It is, therefore, necessary to adopt the nonlinear gyrokinetic simulation model in order to accurately describe the multi-scale dynamics \cite{lee83}.
Up to now, however, electromagnetic gyrokinetic  simulation schemes are based on the nonlinear gyrokinetic equations expressed in terms of potentials \cite{frieman, brizard07}.
Computing such potentials may encounter the so-called cancellation problem \cite{cummings} and/or involve solving, sometimes coupled, Poisson-like equations \cite{manuilskiy, chen03} in multi-dimensional complicated geometries, and thus, could often become extremely intensive and nearly intractable.
That is, while this approach minimizes the mathematical complexity of the underlying equations, it readily yields to demanding exascale-class computing resources. 

%Renewed interest in the gyrokinetic theory arises from the gyrokinetic simulations using electromagnetic fields.

In this work, we adopt a moment equation approach for the fluctuation structures, which is mathematically equivalent to the  conventional nonlinear gyrokinetic simulation model mentioned above. 
In particular, we present a new and novel gyrokinetic simulation model (termed as GK-E\&B); which is formulated based on the nonlinear gyrokinetic equations expressed directly in terms of electric and magnetic fields  \cite{chen19,burby,chen20}. The fields, meanwhile, are solved via momentum balance equations.
Despite the mathematical structure of the equations governing the spatiotemporal evolution of electromagnetic fields is more complicated than the usual Poisson's equations, their properties more readily reflect the plasma response. Thus, we claim that they are in a more suitable form for predictive simulations of long time scale complex plasma behaviors.
To validate this model, we demonstrate that, in uniform plasmas, it can accurately recover,  both analytically and  in numerical simulations, the linear and nonlinear physics of kinetic Alfv\'{e}n waves.

\section{Theoretical Formulation}
Let us consider electromagnetic fluctuations with frequency much lower than the ion cyclotron frequency, i.e., $|\omega|\ll |\Omega_{i}|$, in a  plasma confined by a magnetic field $\mathbf{B}(\mathbf{r},t)$.
The plasma consists of electrons and ions and $m_{e}\ll m_{i}$ with $m_{e}$ and $m_{i}$ being, respectively, electron and ion masses. 
The ions can be of multiple species; e.g., deuterim, tritium and $\alpha$-particles in a fusion plasma.
For low-frequency waves, the wavelengths are, typically, longer than the Debye length and the quasi-neutrality condition, $\sum_{j}q_{j}n_{j}\simeq 0$, is generally valid, with $j$ being the charged particle species. 
Summing up the momentum conservation equations for all the species, we then have
\begin{eqnarray}
\label{eq:momentum}
	\partial_{t} (\rho_{m }\mathbf{U}_{i\perp})=\frac{1}{c}\mathbf{J}\times\mathbf{B}-[\nabla\cdot \mathbf{P}]_{\perp},
\end{eqnarray}
where $\rho_{m}\bm{U}_{i\perp }=\sum_{j\ne e}m_{j} n_{j}\mathbf{U}_{j\perp}$ is the plasma mass flow  due to ions only, $n_{j}=\langle f_{j}\rangle_{v}$, and $\sum_{j\ne e}m_{j} n_{j}=\rho_{m}$ is the ion mass density.
The total stress tensor is defined as $\mathbf{P}=\sum_{j}\mathbf{P}_{j}=\sum_{j}m_{j}\langle \mathbf{v}\mathbf{v}f_{j}\rangle_{v}$, $\langle \cdots\rangle_{v}$ denotes the velocity-space integral, $f_{j}$ is the distribution function of $j$-th  species to be described below, and the subscript $\perp$ refers to the component perpendicular to $\mathbf{B}$.
It is worth mentioning that the stress tensor contains the usual Reynolds stress.
Note that, since $|\mathbf{U}_{e\perp}|\sim|\mathbf{U}_{i\perp}|$, the electron inertia  is negligible in the perpendicular dynamics. Parallel to $\bm{B}$, however, the electron inertia plays important roles in the wave-particle Landau resonance and, thus, needs to be kept in general.
Meanwhile, since light waves are typically suppressed in the low-frequency regime of interest here, the current density $\mathbf{J}$ is given by Amp\.{e}re's law
\begin{eqnarray}
\label{eq:ampere}
	\mathbf{J}=\frac{c}{4\pi}\nabla\times\mathbf{B}.
\end{eqnarray}
It is worthwhile noting that Eq. (\ref{eq:ampere}) is used, here, to compute $\bm{J}(\bm{X},t)$ from $\bm{B}(\bm{X},t)$ and not the other way around, as it is typically done in nonlinear gyrokinetic codes. 

Given $\mathbf{B}(\mathbf{X}, t)$ and $f(\mathbf{X}, \mathbf{v}, t)$, Eq. (\ref{eq:momentum}) can, thus, be used to advance $\rho_{m}\mathbf{U}_{i\perp}$; which, in turn, determines the perpendicular component of the electric field $\mathbf{E}_{\perp}$.
That is, summing up the perpendicular  momentum conservation equation of only the ion species and noting that $|\omega|\ll |\Omega_{j}|$, we have
\begin{eqnarray}
\label{eq:eperp}
	\mathbf{E}_{\perp}=-\frac{1}{c}\mathbf{U}_{i\perp}\times\mathbf{B}+\frac{1}{\rho_{m}}\sum_{j\ne e}\frac{m_{j}}{q_{j}}[\nabla\cdot \mathbf{P}_{j}]_{\perp}.
\end{eqnarray}
%\blue{It is worthwhile noting that Eq. (\ref{eq:eperp}) is formally different from the $\bm{E}_{\perp}$ equation in the gyrokinetic electron and fully kinetic ion particle simulation scheme \cite{chen19},  which resolves the electron dynamics up to $|\Omega_{i}| \ll |\omega|\ll |\Omega_{e}|$.}
Meanwhile, the parallel component of electric field,  $\mathbf{E}_{\parallel}=\mathbf{E}\cdot\mathbf{b}$ with $\mathbf{b}=\mathbf{B}/B$, can be readily obtained by taking the parallel component of  the  time derivative of Amp\.{e}re's law along with the Faraday's law; i.e.,
\begin{eqnarray}
\label{eq:eparallel}
	c^{2}[\nabla_{\perp}^{2} E_{\parallel}-\mathbf{b}\cdot\nabla(\nabla\cdot\mathbf{E}_{\perp})]=4\pi\partial_{t} J_{\parallel}.
\end{eqnarray}
Here, $J_{\parallel}=\sum_{j}J_{j\parallel}$ is the total parallel current and $J_{j\parallel}=q_{j}\langle v_{\parallel}f_{j}\rangle_{v}$. 
Note that the parallel currents are calculated as moments here to account for the particle parallel dynamics.
Once the electric field $\mathbf{E}=\mathbf{E}_{\perp}+E_{\parallel}\mathbf{b}$ is determined, the magnetic field can then be advanced by Faraday's law
\begin{eqnarray}
\label{eq:faraday}
	\partial_{t}\mathbf{B}=-c\nabla\times\mathbf{E}.
\end{eqnarray}
Equations (\ref{eq:momentum}) to (\ref{eq:faraday}) for the electromagnetic fields are closed if we assume that the distribution function for each species is given. 

In the gyrokinetic regime, assuming, for simplicity now, a nearly isotropic plasma, the distribution function $f_{j}$ is given by  \cite{chen19, chen20}, with the subscript $j$ suppressed unless necessary,
\begin{eqnarray}
\label{eq:f}
	f(\mathbf{x},\mu, v_{\parallel},t)= f_{pol} + F,
\end{eqnarray}
where $\mu=v_{\perp}^{2}/(2B)$ is the magnetic moment, $v_{\parallel}$ is the parallel velocity, and 
\begin{eqnarray}
\label{eq:fpol}
	 f_{pol}=\frac{q}{m}[1-T_{g}^{-1}J_{0}]\phi\frac{1}{B}\frac{\partial F}{\partial\mu},
\end{eqnarray}
$T_{g}=\textrm{exp}(-\bm{\rho}\cdot\nabla_{\perp})$ with $\bm{\rho}=\mathbf{b}\times\mathbf{v}/\Omega_{c}$ is the pull-back operator from the gyrocenter coordinate $\mathbf{X}$ to the particle coordinate $\mathbf{x}=\mathbf{X}+\bm{\rho}$, $J_{0}=J_{0}(k_{\perp}\rho)$ is the Bessel function accounting for the finite-Larmor-radius effect, and $k_{\perp}^{2}=-\nabla_{\perp}^{2}$.
In the polarization contribution, Eq. (\ref{eq:fpol}), $\phi$ is defined as $\nabla_{\perp}\phi=-\mathbf{E}_{\perp}$.
Meanwhile, in Eq. (\ref{eq:f}), 
\begin{eqnarray}
\label{eq:F}
	F=T_{g}^{-1}(F_{g})
\end{eqnarray}
is the gyrocenter response and $F_{g}$ satisfies the following nonlinear gyrokinetic equation in its unexpanded form \cite{chen20}
\begin{equation}
\label{eq:fullnlgyrok}
	(\partial_{t}+\dot{\mathbf{X}}\cdot\nabla+\dot{v}_{\parallel}\partial_{v_{\parallel}}) F_{g}(\mathbf{X},\mu,v_{\parallel},t)=0,
\end{equation}
with the gyrocenter phase space motion
\begin{equation}
\label{eq:fullv}
	\dot{\mathbf{X}}=v_{\parallel}\frac{\mathbf{B}_{g}^{*}}{B_{g\parallel}^{*}}+\mathbf{V}_{B}+\mathbf{V}_{E},
\end{equation}
\begin{equation}
\label{eq:fullub}
	\mathbf{V}_{B}=\frac{\mu B}{\Omega B_{g\parallel}^{*}}\mathbf{b}_{g}\times \nabla \langle B_{g}\rangle_{*},
\end{equation}
\begin{equation}
\label{eq:fullue}
	\mathbf{V}_{E}=\frac{c\langle \mathbf{E}_{\perp}\rangle\times\mathbf{b}_{g}}{B_{g\parallel}^{*}}
\end{equation}
and
\begin{equation}
\label{eq:fulldotu}
	\dot{v}_{\parallel}= \frac{\mathbf{B}^{*}_{g}}{B_{g\parallel}^{*}}\cdot[\frac{q}{m}\langle \mathbf{E}\rangle-\mu\nabla\langle B_{g}\rangle_{*}].
\end{equation}
Here, the modified magnetic field has the form
\begin{equation}
\label{eq:bstarfull}
	\mathbf{B}^{*}_{g}=\mathbf{B}_{g}+\frac{v_{\parallel} B}{\Omega}\nabla\times\mathbf{b}_{g},
\end{equation}
$\mathbf{B}_{g}=\langle \mathbf{B}\rangle$ of $\bm{B}$ represents averaging over the gyrophase angle, $\mathbf{b}_{g}=\mathbf{B}_{g}/B_{g}$, $B_{g\parallel}^{*}=\mathbf{B}_{g}^{*}\cdot\mathbf{b}_{g}$, and $\langle \cdots\rangle_{*}$ denotes the gyrophase averaging at an effective Larmor radius of $\rho/\sqrt{2}$ \cite{porazik}.

With the distribution function $f_{j}$, one can then readily calculate $n_{j}$, $\mathbf{P}_{j}$, $\partial_{t} J_{j\parallel}$, and thereby, $\mathbf{E}_{\perp}$ and $E_{\parallel}$.
Taking, as an illustrative example of this approach, the $k_{\perp}^{2}\rho^{2}_{j}\ll 1$ limit in order to simplify the presentations, we have,
\begin{equation}
\label{eq:n}
	n_{j}=n_{pol,j}+N_{j},
\end{equation}
with $n_{pol,j}\simeq -\nabla\cdot [(N_{j}q_{j})/(m_{j}\Omega_{j}^{2})\mathbf{E}_{\perp 0}]$, and $N_{j}=\langle F_{j}\rangle_{v}$. 
Here, from Eq. (\ref{eq:eperp}), we find
\begin{eqnarray}
\label{eq:eperp0}
	\mathbf{E}_{\perp 0}=-\frac{1}{c}\mathbf{U}_{i\perp}\times\mathbf{B}+\frac{1}{\rho_{m}}\sum_{j\ne e}\frac{m_{j}}{q_{j}}[\nabla\cdot \mathbf{P}_{g,j}]_{\perp},
\end{eqnarray}
and $\bm{P}_{g,j}=m_{j}\langle \bm{v}\bm{v}F_{j}\rangle_{v}$.
Meanwhile, we have the stress tensor as
\begin{equation}
\label{eq:p}
	\mathbf{P}_{j}=\mathbf{P}_{pol,j}+\mathbf{P}_{g,j},
\end{equation}
\begin{equation}
\label{eq:ppol}
	\mathbf{P}_{pol,j}\simeq -(3/4)\mathbf{I}\nabla\cdot [(N_{j}q_{j})\rho_{tj}^{2}/2)\mathbf{E}_{\perp 0}], 
\end{equation}
with $\rho_{tj}=v_{tj}/|\Omega_{j}|$ being the thermal gyroradius and $N_{j}v_{tj}^{2}=\langle v_{\perp}^{2}F_{j}\rangle_{v}$.
%Substituting Eqs. (\ref{eq:p}) into Eq.(\ref{eq:eperp}), one obtains the perpendicular electric field $\mathbf{E}_{\perp}$ in the form of a series expansion.
We then have, from Eq. (\ref{eq:eperp}),
\begin{equation}
\label{eq:eperp01}
	\mathbf{E}_{\perp}=\mathbf{E}_{\perp 0}+\mathbf{E}_{\perp 1},
\end{equation}
and 
\begin{equation}
\label{eq:eperp1}
	\mathbf{E}_{\perp 1}\simeq -(3/4\rho_{m})\nabla\sum_{j\ne e}\nabla\cdot [(N_{j}m_{j})\rho_{tj}^{2}/2)\mathbf{E}_{\perp 0}].
\end{equation}
Thus, at least in the $|\rho_{i}^{2}\nabla^{2}_{\perp}|\ll 1$ limit, $\bm{E}_{\perp}$ can be solved algebraically in terms of a series expansion.
%\blue{Within the present approach, the pressure tensor is evaluated to the lowest order in  $\epsilon_{B}\sim |\rho_{i}\nabla\ln B|$, so it can be formally written into the CGL form \cite{chen19} as $\bm{P}_{g}=P_{g\perp}(\bm{I}-\bm{b}\bm{b})+P_{g\parallel}\bm{b}\bm{b}$, where $P_{g\perp}=m\langle \mu B F \rangle_{v}$ and $P_{g\parallel}=m\langle v_{\parallel}^{2} F \rangle_{v}$. }
The term $\partial_{t}J_{\parallel}$, furthermore, is given by taking the $qv_{\parallel}$ moment of the nonlinear gyrokinetic equation, Eq.(\ref{eq:fullnlgyrok}),
\begin{equation}
\label{eq:jparallel}
	\partial_{t}J_{j\parallel}=q_{j}\langle J_{0}[F_{g}\partial(v_{\parallel}\dot{v}_{\parallel})/\partial v_{\parallel}-\dot{\mathbf{X}}\cdot\nabla(v_{\parallel}F_{g})]\rangle_{j, v}.
\end{equation}
Combining Eq. (\ref{eq:jparallel}) and Eq. (\ref{eq:eparallel}) yields the parallel electric field $E_{\parallel}$. 
To be more specific, let us further assume $J_{\parallel}\simeq J_{\parallel e}$ and $|k_{\perp}^{2}\rho_{e}^{2}|\ll 1$. 
Applying Eqs. (\ref{eq:fullv}) and (\ref{eq:fulldotu}) into Eq. (\ref{eq:jparallel}), Eq. (\ref{eq:eparallel}) can then be readily shown to become
\begin{eqnarray}
\label{eq:eparallelnew}
	& &(c^{2}\nabla_{\perp}^{2} -\omega_{pe}^{2})E_{\parallel}\nonumber\\
	&=&c^{2}\mathbf{b}\cdot\nabla(\nabla\cdot\mathbf{E}_{\perp 0})-\frac{4\pi q_{e}}{m_{e}}[(\bm{b}\cdot \nabla)P_{g\parallel}+(P_{g\parallel}-P_{g\perp})\nabla\cdot\bm{b}\nonumber\\
	& &+m\nabla\cdot\langle v_{\parallel}(\bm{V}_{E}+\bm{V}_{B}+\bm{V}_{\kappa})F_{g}\rangle_{v}]_{e},
\end{eqnarray}
where the pressures are given by $P_{g\parallel}=m\langle v_{\parallel}^{2}F_{g}\rangle_{v}$ and $P_{g\perp}=m\langle \mu B F_{g}\rangle_{v}$, $\bm{V}_{B}$ and $\bm{V}_{E}$ are defined, respectively, by Eqs. (\ref{eq:fullub}) and (\ref{eq:fullue}), and $\bm{V}_{\kappa}=(v_{\parallel}^{2}B/\Omega B_{g\parallel}^{*})\nabla\times\bm{b}$ is the magnetic curvature drift.
%Substituting Eqs. (\ref{eq:p}) into Eq.(\ref{eq:eperp}), one obtains the perpendicular electric field $\mathbf{E}_{\perp}$ in the form of a series expansion.
Equation (\ref{eq:eparallelnew}) then yields the parallel electric field $E_{\parallel}$ as
\begin{eqnarray}
\label{eq:eparallel01}
	 E_{\parallel}=E_{\parallel 0}+ E_{\parallel 1},
\end{eqnarray}
where
\begin{eqnarray}
\label{eq:eparallel0}
	E_{\parallel 0}&=&\frac{1}{N q_{e}}[(\bm{b}\cdot \nabla)P_{g\parallel}+(P_{g\parallel}-P_{g\perp})\nabla\cdot\bm{b}\nonumber\\
	& &+m\nabla\cdot\langle v_{\parallel}(\bm{V}_{E}+\bm{V}_{B}+\bm{V}_{\kappa})F_{g}\rangle_{v}]_{e}-d^{2}_{e}\mathbf{b}\cdot\nabla(\nabla\cdot\mathbf{E}_{\perp 0}),
\end{eqnarray}
with $d_{e}^{2}=c^{2}/\omega_{pe}^{2}$ being the electron collisionless skin depth, and $E_{\parallel 1}$ is given by
\begin{eqnarray}
\label{eq:eparallel1}
	(d_{e}^{2}\nabla_{\perp}^{2}-1)E_{\parallel 1}=-d_{e}^{2}\nabla_{\perp}^{2} E_{\parallel 0}.
\end{eqnarray}
Note that, for many applications such as Alfv\'{e}n waves and instabilities, one has $|d_{e}^{2}\nabla_{\perp}^{2}|\sim |k_{\perp}^{2}\rho_{i}^{2}|d_{e}^{2}/\rho_{i}^{2}\simeq |k_{\perp}^{2}\rho_{i}^{2}|(m_{e}/m_{i}\beta_{i})\ll 1$, and, hence, $E_{\parallel 1}\simeq d_{e}^{2}\nabla_{\perp}^{2} E_{\parallel 0}$, and one needs not to solve the Poisson's equation (\ref{eq:eparallelnew}) or (\ref{eq:eparallel1}).
This approximation, however, breaks down for tearing modes, where Eq. (\ref{eq:eparallelnew}) needs to be solved near the singular surfaces where $|\bm{b}\cdot\nabla|$ vanishes. 
Furthermore, in contrast to simulation models which employ the generalized parallel momentum $p_{\parallel}$ variable and potentials in the nonlinear gyrokinetic equations \cite{frieman, brizard07}, our model employs the $v_{\parallel}$ variable as well as the $\bm{E}$ and $\bm{B}$ fields directly. As a consequence, the $E_{\parallel}$ calculation is straightforward and there is no `cancellation' issue \cite{cummings}.
Finally, we note that in the ideal magnetohydrodynamic (MHD) limit, $m_{e}/m_{i}, |d_{e}^{2}\nabla_{\perp}^{2}|, |\rho^{2}\nabla_{\perp}^{2}|\to 0^{+}$, Eq. (\ref{eq:eparallel01}) yields $|E_{\parallel}|\to 0^{+}$, as expected.
 
%\blue{After some straightforward algebra, one can demonstrate that the perpendicular Laplacian term in Eq. (\ref{eq:eparallel}) is negligible for the drift-Alfv\'{e}n waves with $|c^{2}\nabla_{\perp}^{2}/\omega_{pe}^{2}|\sim \mathcal{O}(m_{e}/(\beta_{e}m_{i}))\ll 1$, where $\omega_{pe}$ is the electron plasma frequency.}
%\blue{Furthermore, by solving the Faraday's law for the magnetic field and using gauge-free nonlinear gyrokinetic equations, the conventional Amp\.{e}re cancellation problem, which arises when the parallel canonical momentum $p_{\parallel}$ is used in high $\beta$ simulations \cite{cummings}, has been avoided in the present model.}

\section{Analytical Validation}
As a first step toward demonstrating the validity and usefulness of this new simulation model, we show analytically that it does give correct linear dispersion relation of Kinetic Alfv\'{e}n wave (KAW) in a uniform plasma immersed in a uniform background magnetic field, $\mathbf{B}=B_{0}\hat{z}$.
Let us consider a linear wave with frequency $\omega$ and wave-vector $\mathbf{k}=(k_{\perp}, 0, k_{\parallel})$, and linearize the equations with $\mathbf{U}_{i}=\delta \mathbf{U}_{i}$, $\mathbf{E}=\delta \mathbf{E}=(\delta E_{1}, \delta E_{2},\delta E_{\parallel})$, $\mathbf{B}=\mathbf{B}_{0}+\delta\mathbf{B}$, $\delta \mathbf{B}=(\delta B_{1}, \delta B_{2},\delta B_{\parallel})$ and $F_{g}=F_{g0}+\delta F_{g}$.
Furthermore, we assume $|k_{\parallel}/k_{\perp}|\ll 1$ and $1\gg \beta\gg m_{e}/m_{i}$ with $\beta$ being the ratio between plasma and magnetic pressure. 
For KAWs, we thus have $|\omega|\sim \omega_{A}$ and $|k_{\parallel}v_{te}|\gg |\omega|\gg |k_{\parallel}v_{ti}|$, where $\omega_{A}=|k_{\parallel}|v_{A}$ and $v_{A}=B_{0}/\sqrt{4\pi \rho_{m}}$ are, respectively, the Alfv\'{e}n frequency and speed.
With $|\omega|\ll |k_{\perp}v_{A}|$, the compressional Alfv\'{e}n (fast) wave is, thus, suppressed; that is, $\delta E_{2}\propto \delta B_{\parallel}\simeq 0$. 
It is then straightforward to derive, from Eqs. (\ref{eq:momentum}), (\ref{eq:ampere}), (\ref{eq:eperp}) and (\ref{eq:faraday}),
\begin{eqnarray}
\label{eq:dispersion1}
	& &k_{\parallel}\delta E_{1}[1-\frac{k_{\parallel}^{2}v_{A}^{2}}{\omega^{2}}(1+\frac{ik_{\perp}P_{pol,i}}{n_{0} q_{i}})]\nonumber\\
	&=&-k_{\perp}\delta E_{\parallel}(1+\frac{ik_{\perp}P_{pol,i}}{n_{0} q_{i}})[\frac{k^{2}_{\parallel}v_{A}^{2}}{\omega^{2}}-\frac{ik_{\parallel}P_{i,3}}{n_{0} q_{i}}],
\end{eqnarray}
where the term involving $P_{i,3}\delta E_{\parallel}$ corresponds to $(\mathbf{P_{g,i}})_{1,1}$ due to  $\delta E_{\parallel}$ via $\delta F_{gi}$ given by the linearized ion gyrokinetic equation, Eq. (\ref{eq:fullnlgyrok}).
One readily finds that $|k_{\parallel}P_{i,3}/(n_{0}q_{i})|\sim\mathcal{O}(|k^{2}_{\parallel}v_{ti}^{2}|/\omega^{2})$ and, with $\beta_{i}\ll 1$, is negligible.
Meanwhile, from Eq.(\ref{eq:ppol}), one has
\begin{equation}
\label{eq:ppoli}
	k_{\perp}P_{pol,i}\simeq -\frac{3i}{4}n_{0}q_{i}b_{i}
\end{equation}
with $b_{i}=k_{\perp}^{2}\rho_{i}^{2}/2$. 
Noting that $|k_{\perp}\rho_{e}|^{2}\ll 1$ for KAWs and the small mass ratio  $m_{e}/m_{i}\ll 1$, the parallel current is mainly carried by electrons, i.e., $\partial_{t}J_{\parallel}\simeq \partial_{t}J_{e\parallel}$, and the electron finite Larmor radius effect can be neglected,  i.e., $|f_{pol}|\to 0^{+}$ and $|J_{0}|, |T_{g}|\to 1$ for electrons.
Equation (\ref{eq:eparallel}) along with the linearized Eqs. (\ref{eq:fullnlgyrok}) and (\ref{eq:jparallel}) then yields
\begin{equation}
\label{eq:dispersion2}
	(1+k_{\perp}^{2}d_{e}^{2})\delta E_{\parallel}=i\frac{k_{\parallel}P_{e,3}}{n_{0} q_{e}}\delta E_{\parallel}+k_{\parallel}k_{\perp} d_{e}^{2}\delta E_{1},
\end{equation}
where the collisionless skin depth $d_{e}=c/\omega_{pe}$,
\begin{equation}
\label{eq:pe3}
	k_{\parallel}P_{e,3}\simeq -in_{0}q_{e}[1+2\alpha_{e}^{2}(1-2\alpha_{e}^{2}+i\delta_{e})].
\end{equation}
Here, $\alpha_{e}=\omega/|k_{\parallel}|v_{te}$ and $\delta_{e}=\sqrt{\pi}\alpha_{e}e^{-\alpha_{e}^{2}}$ accounts for the electron Landau damping effect.
%$\Gamma_{p}=1/(1+3b_{i}/4)$, 
Equation (\ref{eq:dispersion2}) then reduces to 
\begin{eqnarray}
\label{eq:dispersion3}
	k_{\perp}\delta E_{\parallel}[\frac{b_{i}m_{e}}{\beta_{i}m_{i}}-\alpha_{e}^{2}(1 -2\alpha_{e}^{2}+i\delta_{e})]=\frac{b_{i}m_{e}}{\beta_{i}m_{i}}k_{\parallel}\delta E_{1},
\end{eqnarray}
where $\tau=T_{e}/T_{i}$.
Combining Eq. (\ref{eq:dispersion1}) with Eq. (\ref{eq:dispersion3}) straightforwardly yields the following KAW dispersion relation
\begin{eqnarray}
\label{eq:dispersion}
	%\frac{\omega^{2}}{\omega_{A}^{2}}=(1+\frac{3}{4}b_{i})[1+\frac{\tau b_{i}}{\frac{\omega^{2}}{\omega_{A}^{2}}(1 -2\alpha_{e}^{2}+i\delta_{e})-\tau b_{i}}],
	\frac{\omega^{2}}{\omega_{A}^{2}}=1+\frac{3}{4}b_{i}+\frac{\tau b_{i}}{1 -2\alpha_{e}^{2}+i\delta_{e}},
\end{eqnarray}
in agreement with the well-known analytical result \cite{hasegawa75, hasegawa76}.
%\begin{equation}
%\label{eq:dispersion4}
	%\frac{\omega^{2}}{\omega_{A}^{2}}=1+b_{i}(\frac{\tau}{1+i\delta_{e}}+\frac{3}{4}).
%\end{equation}

\section{Numerical Simulations}
Here, we demonstrate the validity of this new GK-E\&B simulation model by benchmarking results of a single-wave KAW simulation against the analytical theories in a uniform plasma. Again, we adopt isotropic Maxwellian backgrounds, express $\partial_{t}\delta J_{\parallel}\simeq \partial_{t}\delta J_{e\parallel}$ in terms of the electron parallel momentum balance Eq. (\ref{eq:jparallel}), and, thereby, Eq. (\ref{eq:dispersion2}). We also neglect the compressional component of magnetic field fluctuation in the low-$\beta$ and $|k_{\parallel}/k_{\perp}|\ll 1$ limit.

\begin{figure}[!htp]
\centering
\includegraphics[scale=0.25]{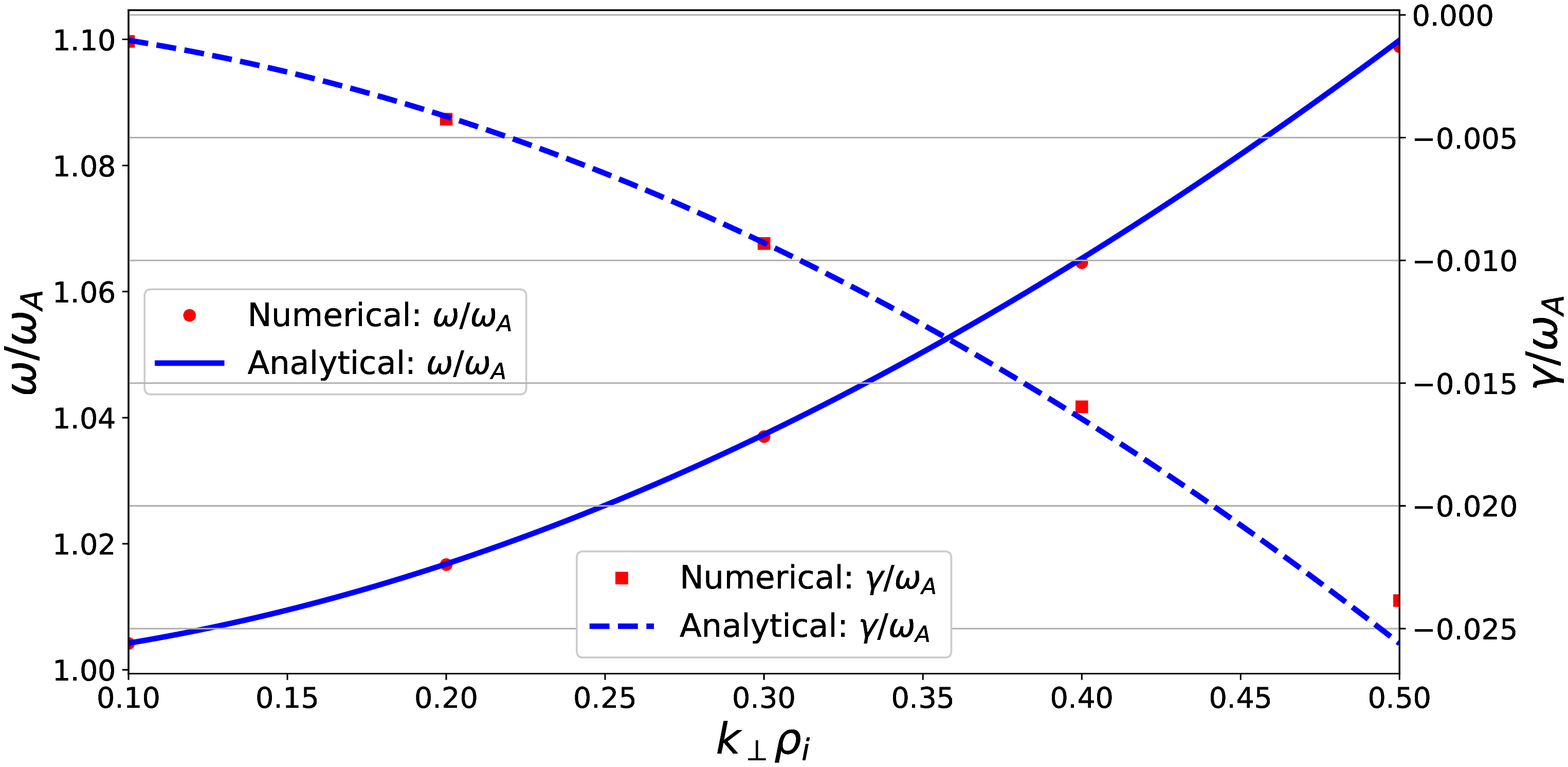}
	\caption{(Color online) Numerical simulation results of KAW complex frequencies versus $k_{\perp}\rho_{i}$ for  $\tau=1$, $\beta_{i}=0.01$, $m_{i}/m_{e}=1836$ and $k_{\perp}/k_{\parallel}=10$.  Lines are solutions of Eq. (\ref{eq:dispersion}).}
\label{eps:dispersion}
\end{figure}

In the linear limit, we can simply assume the perturbed quantities have the form $\delta Q=\delta \hat{Q}e^{i\bm{k}\cdot\bm{x}}+c.c.$, then the governing equations become ordinary differential equations, which can be straightforwardly advanced by the second-order Runge-Kutta scheme with each time step consisting of two sub-steps.
Specifically,  given the fluid variables $\delta\bm{U}^{n}$, $\delta \bm{B}_{\perp}^{n}$, $\delta \bm{P}_{i}^{n}$ and $\delta \bm{P}_{e}^{n}$ at the $n$-th time step, the electric field $\delta \bm{E}^{n}$ is readily obtained from the algebraic equations (\ref{eq:eperp}) and (\ref{eq:dispersion2}). 
The first sub-step calculates kinetic and thus fluid field values at step $n+1/2$ from Eqs. (\ref{eq:momentum}, \ref{eq:faraday}, \ref{eq:fullnlgyrok}), yielding $\delta\bm{U}^{n+1/2}$, $\delta \bm{B}_{\perp}^{n+1/2}$, $\delta \bm{P}_{i}^{n+1/2}$ and $\delta \bm{P}_{e}^{n+1/2}$.
The second sub-step is subsequently carried out in which variables are pushed from $n+1/2$ to $n+1$, using Eqs. (\ref{eq:momentum}, \ref{eq:faraday}, \ref{eq:fullnlgyrok}). 
Figure (\ref{eps:dispersion}) shows our numerical simulation results plotting the KAW frequency and damping rate vs. $k_{\perp}\rho_{i}$. The time step interval is $\Delta t=0.01/\omega_{A}$, which satisfies the Courant condition $|k_{\parallel}|v_{te}\Delta t\lesssim 1$ posed by electron free streaming.
Results from the new simulation model are in good agreement with the analytical theories, i.e., Eq. (\ref{eq:dispersion}).

Another benchmark of the new model is to demonstrate that it produces accurately the nonlinear physics of the kinetic Alfv\'{e}n wave. Taking a single finite-amplitude KAW, the dominant nonlinear physics is the wave trapping of resonant electrons via $P_{e,3}$ in Eq. (\ref{eq:dispersion2}). More specifically, this phenomenon can be observed by retaining the parallel nonlinearity in the electron drift kinetic equation, i.e., Eq. (\ref{eq:fulldotu}). 
Electrons are simulated utilizing the so-called $\delta f$ particle-in-cell (PIC) method \cite{parker93}. Thus we define the electron weight $w=\delta F/g$, and represent the perturbed electron distribution as $\delta F=\sum_{i=1}^{N}w_{i}S(z-z_{i})\delta(v_{\parallel}-v_{\parallel i})\delta(\mu-\mu_{i})$. Here, $N$ is the total number of markers,
$g$ is the numerically loaded and evolved simulation marker distribution, 
and $S(z-z_{i})=S_{0}((z-z_{i})/\Delta z)$ is the marker shape function with $\Delta z$ being the marker size, and $S_{0}(x)=1$ for $|x|<0.5$ and $0$ for $|x|\ge 0.5$.
In this work, markers are loaded according to the background Maxwellian distribution, and the marker size is chosen to equal to the grid size.
%\blue{Note also that, Eq. (\ref{eq:dispersion2}) indicates that there is a cancellation between $\delta E_{\parallel}$ on the left-hand-side and the adiabatic part of $P_{e,3}$. Numerically, this problem can be overcome by adopting the usual split-weight scheme \cite{manuilskiy, chen03}. Detailed analysis of this issue will be addressed in the future publication.}

Letting, furthermore, the dimensionless time $t \omega_{A} \to t$ and performing the normalizations
\begin{eqnarray}
\label{eq:normalization}
	\zeta=|k_{\parallel}|z,\quad \bm{V}=\bm{v} v_{te},\quad \delta e_{\parallel}=\frac{e\delta E_{\parallel}}{|k_{\parallel}| T_{e}},
\end{eqnarray}
the evolution equation of electron weights then becomes
\begin{eqnarray}
\label{eq:weight}
	\dot{w}=-(1-w)\sqrt{\frac{\beta_{e}m_{i}}{m_{e}}}\delta e_{\parallel}v_{\parallel}.
\end{eqnarray}
$w(t)$ can be readily integrated along the marker trajectories in the phase-space, i.e.,
\begin{eqnarray}
\label{eq:dzeta}
	\dot{\zeta}=v_{\parallel}\sqrt{\frac{\beta_{e}m_{i}}{m_{e}}},
\end{eqnarray}
and
\begin{eqnarray}
\label{eq:dvpa}
	\dot{v}_{\parallel}=-\frac{1}{2}\sqrt{\frac{\beta_{e}m_{i}}{m_{e}}}\delta e_{\parallel}.
\end{eqnarray}

%The linear analysis (Fig. (\ref{eps:dispersion})) suggests KAW is near marginal stability in the long-wavelength regime, therefore, the wave trapping effect may be physically interpreted by considering the particle motion within 
Let us consider a KAW with a finite and constant-amplitude parallel electric field, $2\delta \hat{e}_{\parallel}\cos(\zeta-\omega t)$. Equations (\ref{eq:dzeta}) and (\ref{eq:dvpa}) readily show that the phase-space electron motion can be described by a nonlinear pendulum equation
\begin{eqnarray}
\label{eq:wavetrapping}
	\ddot{\Theta}+\frac{\beta_{e}m_{i}}{m_{e}}\delta \hat{e}_{\parallel}\sin(\Theta)=0,
\end{eqnarray}
in the wave moving frame $-\zeta+\omega t=\pi/2-\Theta$. 
Thus, the separatrix width and bounce frequency for resonant electrons trapped by the KAW are, respectively, given by $\Delta v_{\parallel}=4\sqrt{\delta \hat{e}_{\parallel}}$ and $\omega_{B}\simeq \sqrt{\beta_{e}m_{i}\delta \hat{e}_{\parallel}/m_{e}}$.
The self-consistent dynamics can then be divided into two different regimes: (i)  the weakly nonlinear regime, $|\gamma_{l}|\gg|\omega_{B}|$, in which the KAW damps essentially as in the linear regime; and (ii) the strongly nonlinear regime, $|\gamma_{l}|\ll|\omega_{B}|$, in which the resonant electrons execute rapid bounce motion and the KAW experiences negligible damping.

To verify these features, we have carried out a self-consistent nonlinear simulation of a single-wave KAW with the initial amplitude $\delta\hat{e}_{\parallel}\simeq 0.03$ at $k_{\perp}\rho_{i}=0.3$, $\tau=1$, $\beta_{i}=0.01$, $m_{i}/m_{e}=1836$ and $k_{\perp}/k_{\parallel}=10$. 
The time step is $\Delta t=0.01/\omega_{A}$, the grid number is $64$ in a one-dimensional periodic system with the  domain size of $\zeta$ being $2\pi$.  A total of $32800$ marker particles are loaded. 
This condition corresponds to $\Delta v_{\parallel}\simeq 0.69$ and  $\omega_{B}\simeq 0.74 \gg |\gamma_{l}|\sim (10^{-2})$, and, thus, the dynamics is anticipated to be in the strongly nonlinear regime.

\begin{figure}[!htp]
\centering
\includegraphics[scale=0.4]{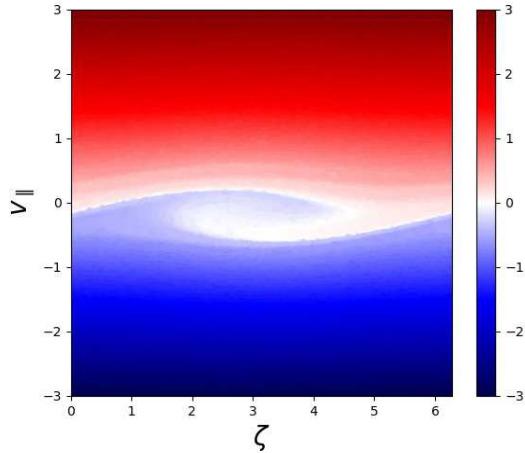}
	\caption{(Color online) Contour of electron marker particles (colored according to the initial value of $v_{\parallel}$) in phase space at $t=15/\omega_{A}$, $k_{\perp}\rho_{i}=0.3$ and $\delta\hat{e}_{\parallel}=0.03$.  The rest of the parameters is the same as Fig. (\ref{eps:dispersion})}
\label{eps:trapping}
\end{figure}

Figure (\ref{eps:trapping}) plots the contour of electron marker particles in the phase space.
It clearly demonstrates wave trapping and the corresponding phase-space structures. The measured separatrix width $\Delta v_{\parallel}\simeq 0.8$ and bouncing frequency $\omega_{B}\simeq 0.8$ agree with the analytical predictions. In addition, not shown here, the wave amplitude remains essentially undamped, as expected theoretically.

\section{Summary and Discussions}
In this work, we have presented a new and novel simulation model, termed as GK-E\&B, for low-frequency electromagnetic waves and instabilities in realistic magnetically confined plasmas.
Contrary to previous hybrid MHD-gyrokinetic simulation models (e.g., \cite{belova,lin01,holod}), this model employs a sophisticated moment approach for electromagnetic fields and  nonlinear gyrokinetic equations using directly electric and magnetic field variables, $\bm{E}$ and $\bm{B}$. 
This approach more readily reflects the plasma response and,  thus, may avoid some of the intensive and complicated computations in previous models using  potentials. More specifically, the new scheme has the advantages that, for practically important applications to Alfv\'{e}n waves and instabilities, the fields could be solved algebraically, and  it intrinsically suffers no conventional Amp\.{e}re cancellation problem.
The current model is also valid for physics from the macroscopic to microscopic scales. 
Thus, kinetic effects such as finite ion Larmor radius and wave-particle interactions are retained.
To demonstrate its validity, we first show analytically that the model reproduces correct linear dispersion relation of the microscopic KAW. We then carry out linear and nonlinear benchmarking simulations, and the results agree well with the analytical predictions. 
While, as noted earlier, we have assumed $k_{\perp}^{2}\rho_{i}^{2}\ll 1$ in the present work in order to simplify the analysis, extending to the regime of arbitrary $|k_{\perp}\rho_{i}|$ is plausible via the Pade's approximation and the results will be reported in the future.

Since low-frequency electromagnetic fluctuations, e.g., Alfv\'{e}n waves and instabilities are prevalent in laboratory and nature plasmas, we believe our GK-E\&B simulation model could provide a powerful tool to extract, understand, and explore the fundamental multi-scale nonlinear processes in a broad scope of magnetized plasmas.
Finally, in order to simplify the analysis and presentation, we have ignored, in the present work, the velocity-anisotropy; which, however, can be readily included following \cite{chen20}.
The application of this new GK- E\&B scheme in more general circumstances, such as more realistic five-dimensional simulations, detailed numerical benchmarks, parameter scans, and a self-consistent treatment of both the nonlinear wave-wave and wave-particle interactions, is also ongoing and will be reported in future publications.
%This and other additional extensions as well as  applications will be reported in a future publication.

\section*{Acknowledgments}
%\label{sec:may:01:16:05}

This work was supported by National Science Foundation of China under Grant Nos. 11235009 and 11905097, and the Fundamental Research Fund for Chinese Central Universities under Grant No. 2019FZA3003.
This work was also carried out within the framework of the EUROfusion Consortium and received funding from the Euratom research and training programme 2014-2018 and 2019-2020 under Grant Agreement No. 633053 (Project No. WP19-ER/ENEA-05). The views and opinions expressed herein do not necessarily reflect those of the European Commission.

\end{document}